\documentclass[a4paper,fleqn,usenatbib]{mnras}  

\usepackage{newtxtext,newtxmath}
\usepackage{graphicx}
\usepackage{geometry}
\usepackage{color}
\usepackage{multicol}
\usepackage[]{hyperref}
\usepackage{amsmath}	
\usepackage{amssymb}	
   \title[GW and EM transient rates]{A consistent estimate for Gravitational Wave and Electromagnetic transient rates}

   \author[Eldridge, Stanway \& Tang]{J. J. Eldridge$^1$
          \thanks{j.eldridge@auckland.ac.nz},
          E. R. Stanway$^2$ and
          Petra N. Tang$^1$
          \\
          $^1$Department of Physics, University of Auckland, Private Bag 92019, Auckland, New Zealand\\
           $^2$Physics Department, University of Warwick, Gibbet Hill Road, Coventry, CV4 7AL, United Kingdom\\
             }

\date{Accepted 2018 October 3. Received 2018 October 2; in original form 2018 July 19}

\pubyear{2018}

\begin{document}
\label{firstpage}
\pagerange{\pageref{firstpage}--\pageref{lastpage}}
\maketitle

  \begin{abstract}
   {Gravitational wave transients, resulting from the merger of two stellar remnants, are now detectable. The properties and rates of these directly relates to the stellar population which gave rise to their progenitors, and thus to other, electromagnetic transients which result from stellar death.}
   {We aim to estimate simultaneously the event rates and delay time distribution of gravitational wave-driven compact object mergers together with the rates of core collapse and thermonuclear supernovae within a single consistent stellar population synthesis paradigm.}
   {We combine event delay-time distributions at different metallicities from the Binary Population and Spectral Synthesis (BPASS) models with an analytic model of the volume-averaged cosmic star formation rate density and chemical evolution to determine the volume-averaged rates of each event rate at the current time.}
   {We estimate rates in excellent agreement with extant observational constraints on core-collapse supernovae, thermonuclear supernovae and long GRBs. We predict rates for gravitational wave mergers based on the same stellar populations, and find rates consistent with current LIGO estimates. We note that tighter constraints on the rates of these events will be required before it is possible to determine their redshift evolution, progenitor metallicity dependence or constrain uncertain aspects of stellar evolution.}
   
   \end{abstract}

   \begin{keywords}
   Methods: numerical -- Gamma-ray burst: general -- supernovae: general -- Gravitational waves
   \end{keywords}


\section{Introduction}\label{sec:intro}

The transient sky is being studied in ever more detail, by a rapidly increasing number of facilities and surveys. These identify astrophysical sources which vary significantly in luminosity on short timescales - typically ranging from minutes to months. While many such variations are associated with recurrent sources, such as cataclysmic variables or accretion onto active galactic nuclei, others are associated with the dramatic energy release events which accompany stellar death.

The dramatic types of stellar death  fall into two broad categories, based on of the driving processes that cause them. Either they are the result of evolutionary processes affecting the core of stars, or they result from orbital evolution as gravitational wave emission carries away energy and angular momentum, driving two compact remnants into contact. The first is best exemplified by the cessation of nuclear burning in an iron-group elements (or oxygen-neon) core of a massive star that then undergoes core collapse resulting in the emission of an optically-luminous transient classified as a supernova (SN). Where hydrogen remains in the stellar envelope these are classified as Type II, while hydrogen-free core-collapse supernovae (CCSNe) are classified as Type Ib or Ic. In a subset of stripped-envelope core-collapse events, a relativistic jet may be launched causing beamed, high-energy emission which may be detected as a Long Gamma-Ray burst (LGRB) by observers on the jet emission axis \citep[e.g][]{2003ApJ...591..288H,2012ARA&A..50..107L,2015PASA...32...16S}.

Spectacularly, the LIGO and VIRGO facilities have now opened a new window onto the transient sky, detecting the gravitational wave (GW) strain variations associated with the second category of transients: the merger of compact binaries. In these events, either two black holes (BH) or neutron stars (NS) lose angular momentum through GW emission and spiral towards a final merger and associated `chirp' event \citep{2016PhRvL.116f1102A,PhysRevLett.119.161101}. While these events had been theorised, and the mechanism of short GRBs (SGRBs) tentatively assigned to GW-driven NS-NS mergers \citep{1986ApJ...308L..43P}, the detection of gravitational wave chirps has finally provided conclusive evidence for this interpretation.

Type Ia SNe may have progenitor pathways that belong in both these categories. White dwarfs, supported by electron degeneracy pressure, form the end state in the life of 95\% of stars. However either accretion onto a white dwarf from a companion or the merger of two white dwarfs in a close system, due to gravitational waves, can cause the star to exceed the Chandrasekhar limit and if the conditions are correct ignite, giving rise to a thermonuclear detonation known as a Type Ia SN \citep[e.g][]{2011NatCo...2E.350H,2014ARA&A..52..107M}.

Current supernova surveys such as PanSTARRS \citep{2016arXiv161205560C}, ATLAS \citep{2018PASP..130f4505T} and ASAS-SN \citep{2017MNRAS.464.2672H} report large samples of supernovae on a weekly basis, while forthcoming surveys, notably those associated with the Large Synoptic Survey Telescope \citep[LSST, ][]{abell2009lsst}, will produce data on thousands such sources per night. Gamma-Ray bursts are rarer, being detected at a rate of a few per week. GW transients are currently the rarest known, although this is likely a result of the enormous technical difficulty of detecting such events rather than their intrinsic rate. 

While the mechanism behind each type of transient, the progenitor stars and the detection methods vary, fundamentally all these events represent a probe of the stellar population which gives rise to them, and our understanding of the stellar evolution processes involved. Their relative rates in a population of known age and metallicity can provide insight into the relative fraction of stars in different mass ranges. As the sample grows over the coming years it will be possible to invert this, by assuming an initial mass function and adopting models for the delay time between stellar birth and transient event, to yield information on the cosmic star formation history and metallicity evolution. 

The strongest such constraints will be obtained from a joint analysis of all available transient samples. However simultaneous modelling of the varying progenitor systems is a challenging task, requiring detailed handling of the physics of stellar evolution across a broad mass range. Crucially, it also requires the modelling of stellar binary interactions, which play a key role in the evolution of nearly all transient progenitors \citep[e.g.][]{1992ApJ...391..246P,1998A&A...333..557D,2000A&A...362.1046L,2008ApJS..174..223B,2010A&A...515A..89M,2011MNRAS.417..408R,2016MNRAS.462.3302E,2016A&A...589A..64M,2017A&A...601A..29Z, 2017MNRAS.472.1593W,2017A&A...606A.136L,2017NewA...51..122L}. Few stellar population synthesis models exist which are capable of a simultaneous, systematic, uniform analysis of the CCSN, GRB, thermonuclear SN and compact merger transient rates arising from a stellar population.

In this paper we utilise the Binary Population and Spectral Synthesis (BPASS) code to present a simultaneous estimate of the transient event rates expected in both gravitational waves and electromagnetic light, arising from the same underlying fundamental stellar population, which reflects the star formation and chemical evolution histories of the Universe on cosmic scales. The paper is structured as follows: in section \ref{sec:method} we describe the stellar population synthesis models used for this analysis and the delay time distribution arising from single-metallicity populations; in section \ref{sec:evolution} we produce a metallicity and star formation weighted model for the volume-averaged stellar population as a function of redshift and
compare this to observational constraints on the rates of different transient types; in section \ref{sec:disc} we discuss the implications of our analysis and prospects for future observatories. Finally in section \ref{sec:conc} we present a summary of our conclusions

Where necessary we assume a standard $\Lambda$CDM cosmology with $\Omega_M$=0.3, $\Omega_\Lambda$=0.7 and $H_0$=100\,$h$\,km\,s$^{-1}$\,Mpc$^{-3}$ with $h$=0.7


\section{Stellar Transients Population Synthesis}\label{sec:method}

\begin{figure*}
\includegraphics[width=0.98\textwidth]{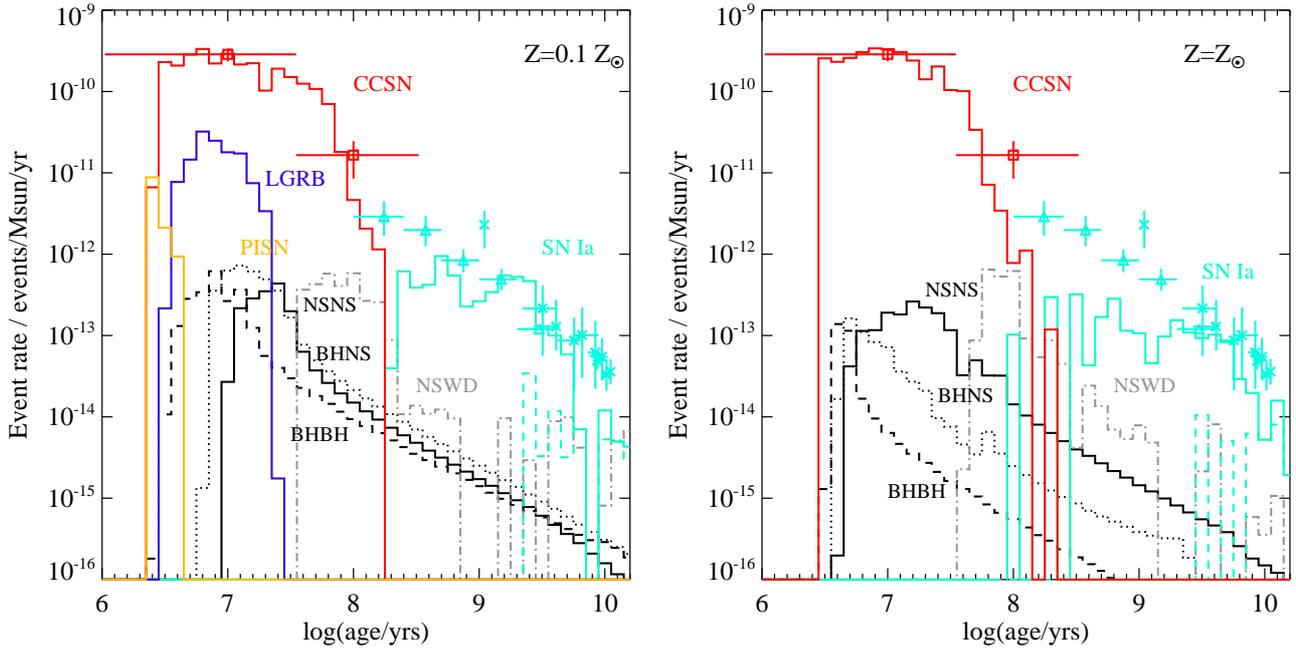}

\caption{The predicted delay time distribution derived from BPASS for electromagnetic and gravitational wave transients at two representative metallicities. 
We compare the predicted delay times at both metallicities with the same set of observational constraints. The square points are the observed core-collapse supernova rates reported by \citet{2010MNRAS.407.1314M}. The remaining data points indicate type Ia SN rates from \citet[crosses]{2010ApJ...722.1879M} and \citet[triangles]{2008PASJ...60.1327T}. The dashed cyan line indicates the delay time distribution for the double degenerate SN Ia pathway, while the solid cyan line is for the all SN Ia.\label{fig:delays}}
\end{figure*}

\subsection{Binary Population and Spectral Synthesis}\label{sec:bpass}

To determine the relative rates of different transient types arising from a stellar population of known age and metallicity we make use of the Binary Population and Spectral Synthesis (BPASS) models\footnote{accessible at bpass.auckland.ac.nz}. These combine a set of custom, detailed, theoretical stellar evolution tracks at thirteen input metallicities. Crucially a substantial subset of the evolution tracks include effects due to binary interactions, with both stars in a binary system of known initial period, mass and mass ratio evolved through detailed modelling in a 1D spherical approximation. Prescriptions are adopted not only for stellar winds but also for mass transfer through Roche Lobe overflow, common envelope evolution, rejuvenation by mass transfer and, more rarely, resulting chemically homogeneous evolution.

The stellar evolution tracks are combined through weighting by an initial mass function (for which we use a broken power-law based on the \citet{2001MNRAS.322..231K} IMF) and by an initial period and mass ratio distribution as a function of primary star mass (for which we use the empirical distributions determined by \citet{2017ApJS..230...15M}). The underlying physical prescriptions and their applications are fully described in \citet{2017PASA...34...58E} and \citet{2018MNRAS.tmp.1296S}. The current version is v2.2, and we present results using our default IMF with a -1.30 slope from 0.1 to 0.5\,M$_\odot$, a Salpeter-like slope of -2.35 above this mass and a maximum mass cut off of 300\,M$_\odot$. We use the 13 metallicities that have been computed to date that range from $Z=10^{-5}$ to $0.040$, and assume $Z_{\odot}=0.020$.

There are two key points to note about binary interaction models in BPASS. First, for common envelope evolution we use our own prescription as described in \citet{2017PASA...34...58E}. It is unique because we use a detailed stellar evolution code so cannot simply remove the hydrogen envelope in one time-step. We therefore remove the envelope as quickly as possible, and also remove its binding energy from the binary orbit. This means we naturally take account of the structure of the star in estimating the outcome of the interaction. Second, during more mild interactions of Roche lobe overflow the companion star is limited in how much it can accrete. For normal stars the limiting factor is the thermal timescale, for white dwarfs and neutron stars we limit the accretion rate by the Eddington luminosity. We do not limit the accretion onto black holes.

BPASS models have previously been used to estimate GW transient properties from simple stellar populations \citep{2016MNRAS.462.3302E} and to explore supernova rates \citep{2015MNRAS.452.2597X} and progenitor populations \citep{2018arXiv180107068X}. They have also been used to evaluate the properties of GRB host galaxies \citep[e.g.][]{2015MNRAS.446.3911S,2017PASA...34...58E}

Stellar death is tracked in a number of ways in the models, which identify core-collapse, the formation of remnants and the further evolution of the secondary star beyond the death of the primary. Many remnants are formed with natal kicks as a result of impulse from a supernova. We adopt a prescription for kicks as described in \citet{2005MNRAS.360..974H}. In brief, we use a Maxwell-Boltzmann distribution with a velocity of 265${\rm km \,s^{-1}}$, sampling this distribution through repeated model iterations to understand the full range of possible outcomes from each SNe. We do not currently employ the kick of \citet{2016MNRAS.461.3747B} or \citet{2018arXiv180404414B}. This kick will only have an impact on our predicted GW event merger rates as shown in \citet{2018arXiv180404414B}, and not on supernova rates.

We discuss the individual key pathways for stellar transients below.

\subsubsection{Core Collapse}

We identify a core-collapse event if central carbon burning has taken place and the CO core mass exceeds 1.38\,M$_{\odot}$ while the total stellar mass is greater than 1.5\,M$_{\odot}$. We classify the event type as described in \citet{2011MNRAS.414.3501E,2013MNRAS.436..774E} and \citet{2017PASA...34...58E}. In brief, an event is classified as type II if the mass of Hydrogen is $>10^{-3}$\,M$_{\odot}$, and Type Ib or c otherwise.

The remnant mass is determined by calculating the residual mass bound to the star after ejection of material given a typical supernova energy injection of $10^{51}$~ergs. This method was first outlined in \citet{2004MNRAS.353...87E} which provides more details. Core-collapse events which produce a remnant mass exceeding 3\,M$_\odot$, and in which the progenitor has experienced chemically homogeneous evolution due to mass transfer at very low metallicity  ($Z\le0.004$), are deemed to explode as long gamma-ray bursts \citep[see][for details]{2017PASA...34...58E}. While this is a probable LGRB progenitor pathway, there are likely others which are not identified within BPASS. In particular, this mechanism is unable to produce progenitors in stellar populations with a metallicity above 0.2\,Z$_\odot$, while such events are known to occur within the observed sample. We intend to explore other pathways in future but improved treatment of stellar rotation, and possibly tidal forces, will be required. The current BPASS LGRB rate estimates should therefore be treated as lower limits.

Finally if the helium core mass at the end of evolution is between 64 and 133\,M\,$_{\odot}$ a pair-instability supernova (PISN) is assumed to occur which completely disrupts the star and leaves no remnant \citep{2002ApJ...567..532H}.

%

\subsubsection{Compact Binary Coalescence}

When the second star in a binary system forms a compact remnant, the orbital evolution of the resultant compact binary is presumed to be dominated by gravitational wave radiation. For each compact binary formed we identify the total delay time, which incorporates both stellar lifetime and the timescale for merger given the remnant masses and the binary period at compact object formation. We adopt the \citet{1964PhRv..136.1224P} prescription for evolution of binary orbits due to gravitational wave emission.

As in \citet{2016MNRAS.462.3302E} we use the near-circular and highly eccentric approximate formulae from \citet{1964PhRv..136.1224P} and interpolate between these for intermediate eccentricities. We note that we may be underestimating the eccentricity of the initial binary because after the first SN in a binary we assume the orbit is circular. 

We calculate rates separately for WD-WD, NS-WD, NS-NS,  NS-BH and BH-BH compact binaries. All of these are technically sources of gravitational waves, although for WD-WD and NS-WD systems the frequency range and signal strength puts mergers below the sensitivity threshold of current and planned detectors. WD-WD mergers are considered type Ia supernovae as described below. The remaining combinations are considered GW transient sources. NS-NS, and perhaps NS-BH, systems are also likely to be electromagnetic transients, with on-axis jetted emission giving rise to a short GRB and more isotropic emission in the form of a fainter kilonova \citep{2017ApJ...848L..12A,2013Natur.500..547T}.

\subsubsection{Type Ia (Thermonuclear) events}

We identify Type Ia thermonuclear supernovae which occur through two complementary channels, both of which result from binary interactions. In the single-degenerate channel a white dwarf accretes material from a main-sequence, brown dwarf or giant companion and detonates when it reaches the Chandrasekhar limit. Here we assume that if a white dwarf with a mass less than 1.2$M_{\odot}$ accretes material and reaches a mass of 1.4$M_{\odot}$ the star will explode as a type Ia SN.

The double-degenerate channel causes a type Ia supernova when two white dwarfs evolve and remain bound in a binary system. Gravitational wave radiation is then assumed to dominate the dynamical evolution of the binary and the merger timescale is calculated using the same method as for NS-NS, NS-BH and BH-BH binaries but assuming a circular orbit for the binary.

Our rates are dominated by the single-degenerate channel. Other channels are hypothesised for subsets of type Ia events, which result in partial detonations \citep[e.g. the sub-luminous events which give rise to hypervelocity white dwarfs, ][]{2018ApJ...858....3R}. However these represent a small fraction of the observed population. 

One key aspect of modelling type Ia SNe from the single-degenerate channel may explain several of these minor pathways:
the accretion rate of material onto the white dwarf \citep[e.g][]{1991ApJ...367L..19N,2000A&A...362.1046L,2004ApJ...613L.129K,2009ApJ...699.2026R,2011MNRAS.417..408R,2017MNRAS.472.1593W}. If the rate is too low then material builds up on the surface of the white dwarf until it ignites in a nova. If it is too high then the Eddington luminosity can be exceeded or a hydrogen envelope can be formed around the white dwarf returning it to the giant branch. There is however a narrow accretion rate range where the material can burn at the surface of the white dwarf as it is accreted, thus increasing the mass of the white dwarf. This mass transfer rate is estimated as around a few $\times \,10^{-7}M_{\odot}\,{\rm yr^{-1}}$ \citep[e.g][]{2011MNRAS.417..408R}. The average accretion rate in our models from start of mass transfer to explosion cover a range of mass transfer rates from $10^{-9}$ to a few $\times \,10^{-5}M_{\odot}\,{\rm yr^{-1}}$. However the models that contribute most of the weight to our Ia rates lie within the narrower range of $10^{-7}$ to a few $\times \,10^{-5}M_{\odot}\,{\rm yr^{-1}}$.

While we still stress that our predicted type Ia rates should be used with caution, the fact they lie within the expected accretion rate window for white dwarf growth, as well as matching observational constraints as shown in this article, suggest that they  represent a fair estimate. Further analysis of the progenitor systems is beyond the scope of this paper.
The improvement in our low mass stellar grid and prescriptions in the current BPASS v2.2 (described in Stanway et al. 2018) included the calculation of a significantly larger grid of stellar models which produce WD binaries, while the revised binary parameter distributions also adopted in v2.2 have further improved our estimates for type Ia rates, which we now consider a robust output of the BPASS code.

\subsection{Delay Time distribution}

A key prediction of any stellar population synthesis model is the distribution in delay time, i.e. the total time required from the onset of star formation to allow stellar evolution to occur and a transient to be triggered.  This distribution will differ by transient type and also with metallicity and as a function of binary interaction parameters.

In figure \ref{fig:delays} we show the delay time distributions predicted by BPASS for electromagnetic and gravitational wave transients at two representative metallicities (Z$_\odot$ and 0.1\,Z$_\odot$). We compare the predicted delay times at both metallicities with the same set of local Universe observational constraints (for which detailed metallicity information on progenitors is unavailable). The square points are the observed core-collapse supernova rates  reported by \citet{2010MNRAS.407.1314M} as a function of host stellar population age. The remaining data points indicate type Ia SN rates from \citet[crosses]{2010ApJ...722.1879M} and \citet[squares]{2008PASJ...60.1327T}.

Core collapse supernovae occur in stellar population ages as high as $\sim200$\,Myr in a binary evolution model, as opposed to the few $\times$10\,Myr limit in single star evolution models. This is a consistent prediction between binary population synthesis models as shown by \citet{2017A&A...601A..29Z}. Furthermore as would be expected, the sources requiring more massive progenitors (LGRBs and PISNe) are restricted to younger ages and lower metallicities.

Thermonuclear supernovae begin to appear in a population at ages of $\sim100$\,Myr, but extend up to (and indeed beyond) the age of the Universe.
For both core collapse and type Ia events, the distribution of observed rates in the local Universe requires the presence of binary pathways in the transient progenitors, and is in good agreement with BPASS predictions. The high rate of observed type Ia events at short timescales of a few 100\,Myr also marginally favours a sub-Solar mean metallicity for their progenitors. The double degenerate pathway for these supernovae only contribute to the rate at ages beyond a few Gyr.  At ages close to 10\,Gyr the double degenerate rate becomes the dominant contribution to thermonuclear events.

By contrast the delay time distribution in gravitational wave transients is considerably broader, with the most rapid, prompt events occurring within a few Myr of the formation of their progenitor stars, and the distribution extending to $>10$\,Gyr. We note that while event rate per year is low at late times for a given initial mass of star formation, the large width of these time bins means that they actually dominate the total number of events from the population.



\section{Cosmic Evolution of Transient Rates}\label{sec:evolution}

\subsection{Input Cosmic History}

The star formation history and chemical evolution of individual galaxies are sensitive to their merger trees and environment. A detailed estimate of transient rates in a given galaxy, such as the Milky Way, must account for this detailed evolutionary history. However on sufficiently large scales, such that both dense and sparse regions of the large scale structure are well sampled, these variations can be averaged out to determine a smooth evolution. \citet{2014ARA&A..52..415M} considered a compilation of galaxy survey data to determine analytic forms for the volume-averaged evolution in both star formation rate density (SFRD) and metallicity mass fraction (Z) over cosmic time. 

Here we use the analytic expressions for SFRD, $\psi(z)$, from \citet{2014ARA&A..52..415M} to follow star-formation through cosmic history. This is given by,
\begin{equation}
\psi(z) = 0.015 \frac{(1+z)^{2.7}}{1+((1+z)/2.9)^{5.6}} \, M_{\odot}\, {\rm yr^{-1} \, Mpc^{-3}}.
\end{equation}
We combine this with the analytic expressions for metallicity evolution used by \citet{2006ApJ...638L..63L}. Here the fractional mass density of star formation at and below metallicity mass fraction of $Z$ is given by,
\begin{equation}
\Psi\left(z,\frac{Z}{Z_{\odot}}\right)=\psi(z) \frac{\hat{\Gamma}[0.84,(Z/Z_{\odot})^2 \, 10^{0.3 z}]}{\Gamma(0.84)},
\end{equation}
where $\hat{\Gamma}$ and $\Gamma$ are the incomplete and complete Gamma functions and we have used the fiducial values for the model from \citet{2006ApJ...638L..63L}. We use this parameterization because it provides a distribution of metallicity at each redshift, rather than other methods requiring a spread around the mean of a metallicity. 

To calculate the the rate of transients at each redshift step, $z_{obs}$, we first calculate the star-formation density at each redshift bin. We then split this star formation between each of our input metallicities. We also calculate the time interval represented by each redshift bin, as well as the age of the Universe in each redshift bin \citep[as in][]{1999astro.ph..5116H}. We then convolve this with the look back time distributions of each transient type at each metallicity, starting at each redshift bin and working back through time to work out the contribution to the current redshift bin from all previous star-formation. 

From the resultant, mixed metallicity, mixed age population, we calculate the volume-averaged mean rate of each transient type, $T$, as a function of redshift:

\begin{equation}
\mathrm{R}(T, z_{obs}) = \sum^{100}_{z_i=z_{obs}} \sum_{Z_i} \mathrm{R}( T, a_i, Z_i ) \times \Psi(z_i,{Z_i}) \times \Delta t(z_i)\,
\end{equation}
where $Z_i$ represents the metallicity distribution at $z_i$ and R($T, z_{obs}$) is the final event rate per year per ($h/0.7$)$^3$\,Gpc$^3$.

\begin{figure}
\includegraphics[width=0.48\textwidth]{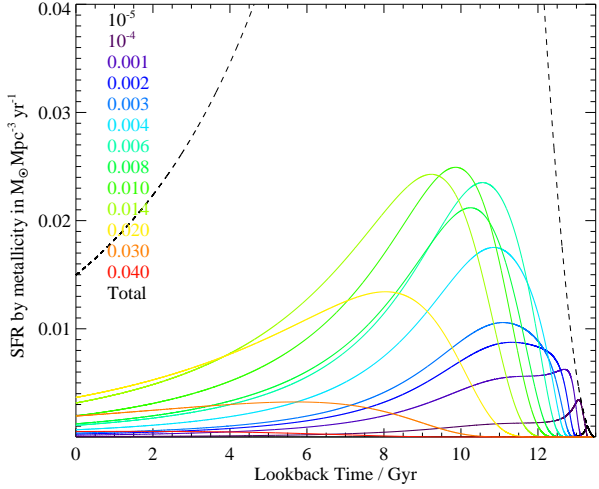}
\caption{The input cosmic star-formation rate and metallicity evolution used to calculate our rates through cosmic history. The dashed line represents the total star-formation rate of all metallicities combined. Each line is colour coded to a specific metallicity\label{metallicity}.}
\end{figure}

We show the resulting star-formation rates in each of our metallicities versus look-back time in Figure \ref{metallicity}. As we can see today the higher metallicities dominate current star formation. At the peak of the star-formation rate history in the Universe, the metallicity is dominated by the approximately half-solar metallicities. Only at the earliest times do the lowest metallicities dominate the star-formation.

\subsection{Rate evolution}

\begin{figure*}
\includegraphics[width=\textwidth]{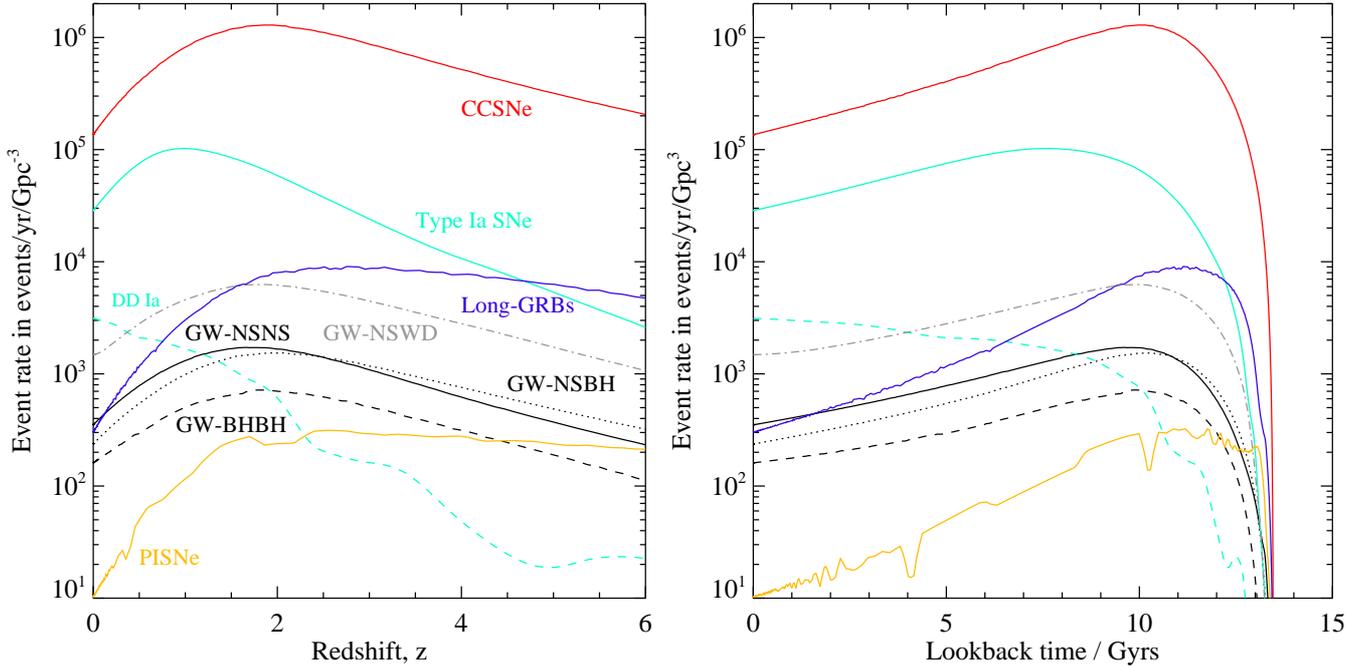}
\caption{The predicted rate density for electromagnetic and gravitational wave transients, as a function of source redshift and lookback time, given the volume-averaged history of cosmic star formation and chemical enrichment. We show both the total SN Ia rate and that due to the double degenerate pathway (DD) alone. \label{fig:rates}}
\end{figure*}

Given this mixed stellar population, we present the rate evolution of each transient type against redshift and lookback time in figure \ref{fig:rates}. 

Core collapse SNe, which arise from massive stars with short delay timescales and which show a relatively weak metallicity dependence, closely track the volume-averaged mean star formation rate. Their volume-averaged population peaks just below $z\sim2$, an epoch known as Cosmic Noon due to the intense galaxy merger, AGN and star formation activity at this redshift.

Type Ia SNe are the second most frequent event type at most redshifts. The long delay time associated with these events (see figure \ref{fig:delays}) shifts the peak in event rate to lower redshifts, $z\sim1$, below which the rate declines. At $z=0$, the volume-averaged type Ia rate is 21\% of that for CCSNe. At the current time the type Ia rate is dominated by the single-degenerate channel. The rate of the much slower double-degenerate channel is still rising and it will dominate the rate in the future.

By contrast, the strong metallicity dependence of long-GRB progenitors in our model shifts the peak in GRB emission rate to higher redshifts, with a near-constant volume-averaged rate at $z\sim2-4$. The inclusion of higher metallicity GRB pathways may modify this redshift evolution in future, but will have little effect on the shape of the evolution at low redshift.

Pair-instability supernovae are very rare events in our population synthesis, and are also heavily biased towards very low metallicities and only a few stellar models in our grid, resulting in a predicted rate that is highly stochastic with redshift. We show the predicted rates smoothed over three time bins. The population peaks at $z\sim3$ in our model, but at a volume-averaged rate more than 1\,dex below that of LGRBs.

The delay times for gravitational wave transients involving black holes and neutron stars are shorter than those for typical white dwarf mergers, but still substantial. As a result, the number density in predicted rate peaks between $z=1$ and $z=2$ for all three flavours of GW event considered here. Low metallicities (i.e. higher formation redshifts) favour the retention of mass during evolution of a stellar progenitor and thus the formation of a black hole, and this manifests as a slightly more rapid fall off in NS-BH events than NS-NS events towards $z=0$.

\subsubsection{Observational Constraints}

\begin{figure}
\includegraphics[width=0.5\textwidth]{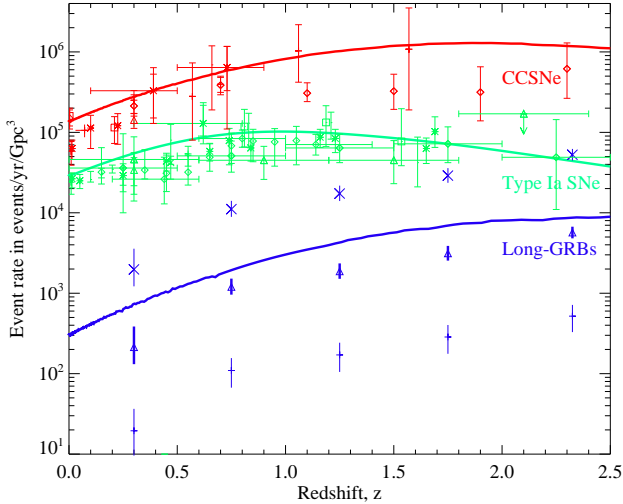}
\caption{The predicted evolution in volume-averaged event rates predicted by BPASS (solid lines) and derived from a compilation of observational data for CCSNe, Type Ia SNe and LGRBs (see text for references). LGRB rates have been corrected to isotropic rates assuming jet opening angles of 8\degr\ (triangles) or 12\degr\ (points), and integrated over the luminosity function for a broader 20\degr\ opening angle (crosses). \label{fig:rates2}}
\end{figure}

In figure \ref{fig:rates2} we compare the predictions with a compilation of observational constraints from a range of surveys and data sources. For observed SN Ia rates we use the data compilation of \citet{2012A&A...545A..96M} and supplement this with rates more recently reported by \citet{2014PASJ...66...49O}, \citet{2015A&A...584A..62C}, \citet{2014AJ....148...13R} and \citet{2014ApJ...783...28G}.  For CCSNe we again use the compilation of \citet{2012A&A...545A..96M} and additionally show the rates reported by \citet{2014ApJ...792..135T}, \citet{2015ApJ...813...93S} and \citet{2016A&A...594A..54P}.

Broadly speaking the agreement between observational constraints and BPASS predictions as a function of redshift is good. For type Ia SNe, the predicted rates consistently tend towards the upper end of the range of observed values, although they coincide with a reasonable fraction of the observed data points, and reproduce the observed redshift trend. A slight overestimate may arise because we have too loose a definition of what leads to a type Ia SN and some fraction of our progenitor models may in fact give rise to some other kind of low luminosity transient as discussed earlier. Type Ia rates are known to differ in cluster environments relative to the field, so a certain amount of scatter is expected. It is also, of course, possible that there are completeness corrections or other systematic offsets in some of the observational data, but overall the predictions, based purely on stellar population synthesis and the volume-averaged cosmic history are excellent. 

The predictions also agree well with observed rates for core collapse supernovae as a function of redshift. The high redshift core-collapse supernova rates reported by Strolger et al (2015, red diamonds on figure \ref{fig:rates2}) are systematically lower than other estimates, and than the BPASS predictions. These are intrinsically challenging deep field observations and the analysis from the CLASH and CANDELS fields necessarily surveys a smaller volume than most other supernova surveys. As a result, these rates may be subject to uncertainties due to cosmic variance. Nonetheless, if these lower CCSN rates are supported by future observations, the implication is that a smaller fraction of core collapse events are resulting in a luminous transient than expected - perhaps supporting the hypothesis that some black-hole forming events are dark \citep{2017MNRAS.468.4968A}.

The volumetric rate of long-GRBs is extremely hard to derive from existing data sets, given the sensitivity to the energy bands and trigger levels in the observed events, together with the extremely non-uniform follow-up and redshift determination for these sources. Further complication is added by the fact that these events are relativistically-beamed and any comparison to theoretical rates requires an estimate of the jet opening angle. In figure \ref{fig:rates2} we show the volumetric rates of events with an isotropic-equivalent energy E$_{iso}>10^{51}$\,ergs reported by \citet{2016ApJ...817....7P} as a function of redshift, given a range of assumptions. In the first case (triangles), we assume that the correction factor for bursts which have jets points pointing away from Earth and are therefore missed from the observed rates is purely geometric (i.e. scales with the jet half-opening angle $\theta$ as [1-cos($\theta$)]$^{-1}$, giving a correction factor of 821 for an opening angle of 8$\deg$). In the second case (points) we instead correct by a factor of $75\pm25$, as calculated by \citet{2005ApJ...619..412G}, a calculation designed to account for the much broader opening angles inferred for low luminosity events. We note that given the redshift-dependence of the GRB luminosity function, the minimum observed luminosity and geometric effects, this likely represents a minimum correction on the observed rates, with larger corrections required at higher redshifts.

Interestingly, despite considering only one plausible pathway for the formation of Long GRBs, and likely missing others, the BPASS rate predictions are higher than those determined by observations by a factor of $\sim2-3$. One possible interpretation is that a large fraction of the events we identify as potential GRBs are in fact stifled at birth, either through inability of the jet to emerge from the stellar photosphere, or through rapid collapse to a black hole from which afterglow radiation cannot escale. We note however that the \citet{2016ApJ...817....7P} study only estimated the rate for the most luminous GRBs. The energy range of observed bursts extends well below the cut-off used by the Perley et al study, and is parameterised by a broken power-law with slope -1.2 below a break luminosity of $5\times10^{50}$\,ergs and -1.92 above this \citep{2016A&A...587A..40P}. Given this luminosity function, we would expect only a few percent of the total population of Long GRBs to be included in the Perley et al (2016) rates, even after the corrections applied by those authors to obtain an estimated rate for E$_{iso}>10^{51}$\,ergs. To illustrate this, we also show the Perley et al rates adjusted to account for the luminosity function integrated over the range $10^{49}-10^{54}$\,ergs, and apply a simple geometric jet correction factor for the somewhat broader jet opening angle of 20\,degrees (crosses). These exceed the rate estimates from BPASS by a factor of $\sim5$, confirming that there is scope for additional evolution pathways to be identified.

For reference, we provide the inferred event rates at $z=0$ and $z=0.5$, accounting for the global volume-averaged star formation and chemical enrichment history, in table \ref{tab:rates}. Our fiducial model is also available for download from the BPASS websites. 

 \begin{table}
 \caption[]{Local Universe transient event rates, given the volume-averaged history of cosmic star formation and chemical enrichment. Quoted uncertainties give Poisson uncertainties on observed predicted event rate based on our model grid, but do not account for statistical offsets or missing pathways.}
    \label{tab:rates}
    \begin{tabular}{lcc}
     \hline
      \noalign{\smallskip}
      Transient Type    &  log$_{10}$\,R($z=0$)$^a$  & log$_{10}$\,R($z=0.5$)$^a$  \\
      \noalign{\smallskip}
      \hline
      \noalign{\smallskip}
      CCSNe   &  5.118 $\pm$ 0.001   & 5.600 $\pm$ 0.001\\
      Ia      &  4.456 $\pm$ 0.003   & 4.878 $\pm$ 0.002 \\
      LGRB    &  2.47  $\pm$  0.02   & 3.07 $\pm$  0.01 \\
      PISN    &  1.01  $\pm$  0.12   & 1.63 $\pm$  0.06 \\
      NSWD    &  3.17 $\pm$  0.01   & 3.45 $\pm$  0.01 \\
      NSNS    &  2.55  $\pm$  0.02   & 2.90 $\pm$  0.02\\
      NSBH    &  2.37  $\pm$  0.03   & 2.74 $\pm$  0.02\\
      BHBH    &  2.21  $\pm$  0.03   & 2.47 $\pm$  0.02\\

            \hline
          \end{tabular}\\

 { $^a$ Rates are given per type in units of events yr$^{-1}$\,$h_{0.7}^{3}$\,Gpc$^{-3}$}
  \end{table}

\begin{figure}
\includegraphics[width=0.5\textwidth]{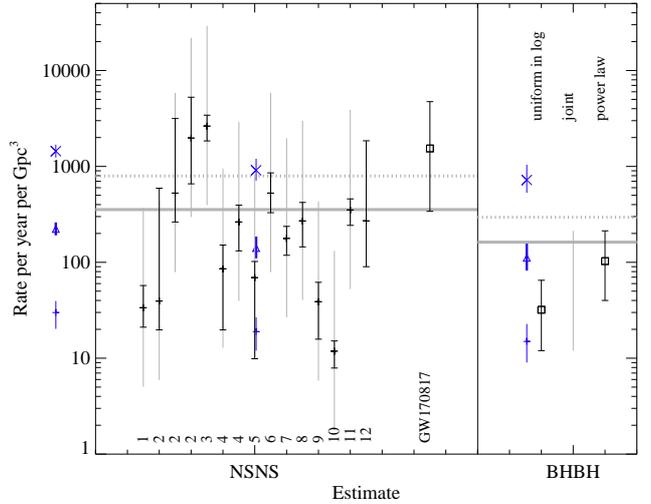}
\caption{The predicted BPASS merger rates of NSNS and BHBH binaries at $z=0$ (horizontal solid line), and $z=0.5$ (dotted line), compared to local rate estimates of events in the literature. Points indicate constraints from short GRBs, while boxes indicate constraints from gravitational wave transient detections. For short GRBs, the extended grey bar indicates the range of number densities assuming jet opening angles $\theta=3-26$\degr\ while bold points indicate the number density and its uncertainty inferred at $\theta=10$\degr, except in the case of ref.\,12 where the uncertainty in opening angles is already accounted for in the error bars shown. Data sources: GW170817 - \citet{PhysRevLett.119.161101}, BHBH rate estimates - \citet{2017PhRvL.118v1101A}, SGRBs: 1 - \citet{2004JCAP...06..007A}, 2 - \citet{2006A&A...453..823G}, 3 - \citet{2006ApJ...650..281N}, 4 - \citet{2009A&A...498..329G}, 5 - \citet{2011A&A...529A..97D}, 6 - \citet{2012MNRAS.425.2668C}, 7 - \citet{2014MNRAS.437..649S}, 8 - \citet{2015MNRAS.448.3026W},  9 - \citet{2004ApJ...609..935Y}, 10 - \citet{2016A&A...594A..84G}, 11 - \citet{2018MNRAS.477.4275P}, 12 - \citet{2015ApJ...815..102F}.  \label{fig:mergerrates}}
\end{figure}

Observational constraints on the local rate of NS-NS and BH-BH mergers are shown in figure \ref{fig:mergerrates}. The majority of constraints on NS-NS mergers arise from the short gamma-ray burst population. As was the case for LGRBs, the conversion from an observed rate to an intrinsic rate density involves an uncertainty on the opening angle assumed for the relativistically-beamed emission. We show the volumetric rate density and its uncertainty assuming a jet opening angle of 10\degr\ and also the range of possible densities for angles in the range 3 to 26\degr\ \citep[after ][]{2018MNRAS.477.4275P}. This also encompasses the opening angle estimate of \citet[][$\theta=16\pm10$\degr]{2015ApJ...815..102F}
We also show the astrophysical rates inferred from the detection of gravitational wave transient GW\,170817 \citet{PhysRevLett.119.161101} and the handful of BH-BH mergers identified to date \citet{2017PhRvL.118v1101A,2017PhRvL.119n1101A,2017ApJ...851L..35A}. 

As figures  \ref{fig:rates2} and \ref{fig:mergerrates} demonstrate, the same stellar population synthesis model which successfully reproduces the rate of core collapse and type Ia supernovae also provides a good fit to the observed rate of binary NS and BH mergers. Although we note our predicted BH merger rates are towards the upper end of the current observational constraints.

No known constraints on the gravitational-wave driven merger of NS-WD binaries exists. These events have been proposed as one possible progenitor for the Ca-rich class of transients \citep{2012MNRAS.419..827M}. The merger rates predicted for NS-WD events by BPASS is lower than current best estimates of the local volumetric rate extimated for Ca-rich SNe \citep[in our units, log(events/\,yr$^{-1}$\,Gpc$^{-3}$) = $4.08^{+0.29}_{-0.17}$, ][]{2018ApJ...858...50F} suggesting that either this progenitor pathway contributes only a fraction of observed events, or that some alternate mechanism is required for hardening NS-WD binaries before the GW-dominated inspiral phase. Other studies involving binary population synthesis have rates similar to or higher than our rate \citep[e.g.][]{2017MNRAS.467.3556B,2018arXiv180401538T}. 

We finally note that our predicted rates are dependent on the assumed star-formation history and metallicity evolution of the Universe. For the star-formation history, the total star formation that occurred at the highest redshifts, beyond $z\sim 5$, is the most uncertain. This only has an impact on the rate of those events with the longest delay times, i.e. the GW events and double-degenerate type Ia SNe. The uncertainties in the cosmic metallicity evolution has the strongest effect on the events with the strongest dependence on metallicity, i.e. LGRBs and PISNe. Here their rate evolution depends strongly on how quickly the Universe became metal-rich. However, considering the good qualitative agreement between our current predictions and observational constraints there is no evidence that the fiducial model for this metallicity and star-formation history should be changed. 

\subsection{Chirp mass distributions of mergers}

\begin{figure}
\includegraphics[width=0.5\textwidth]{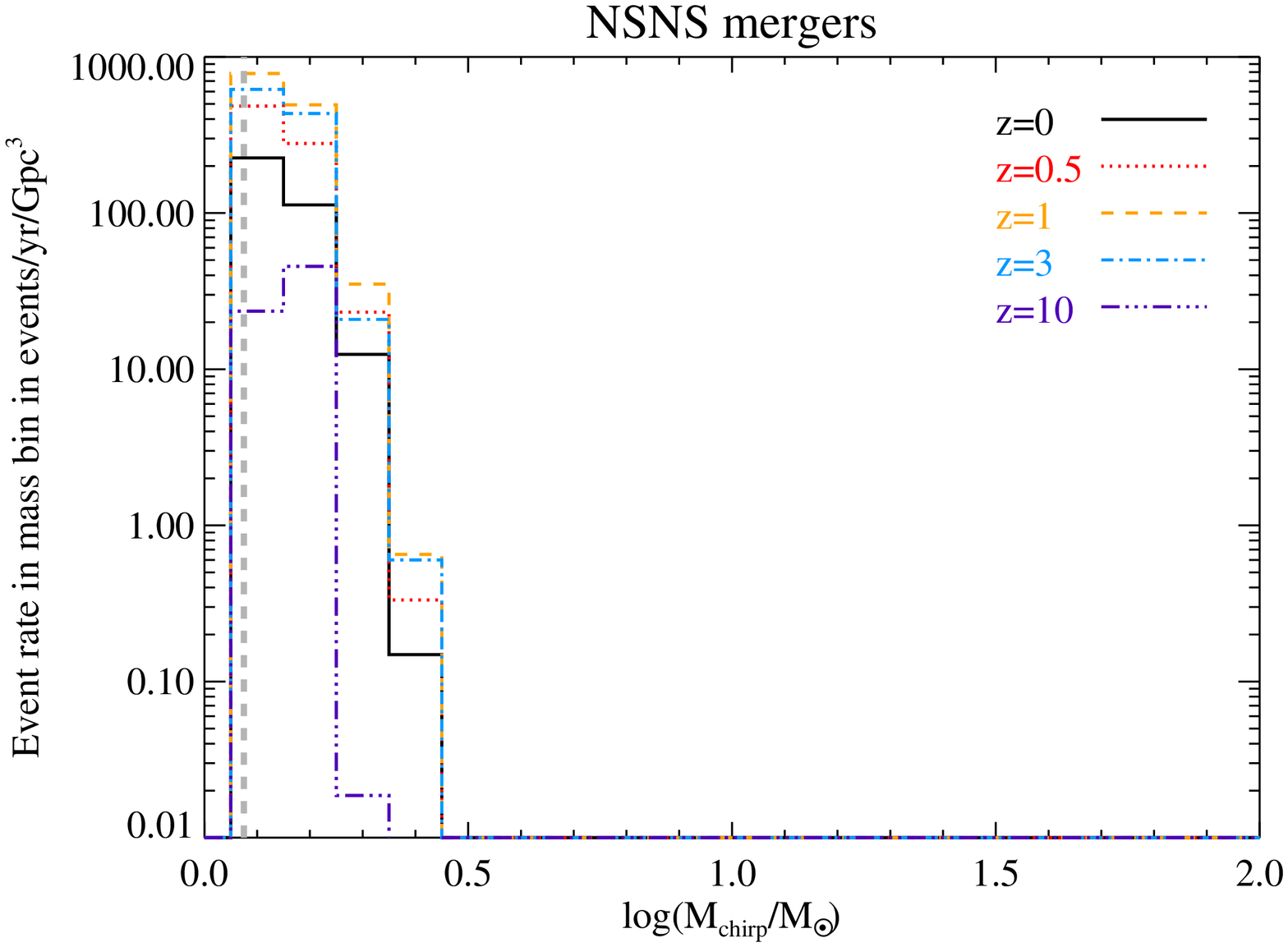}
\includegraphics[width=0.5\textwidth]{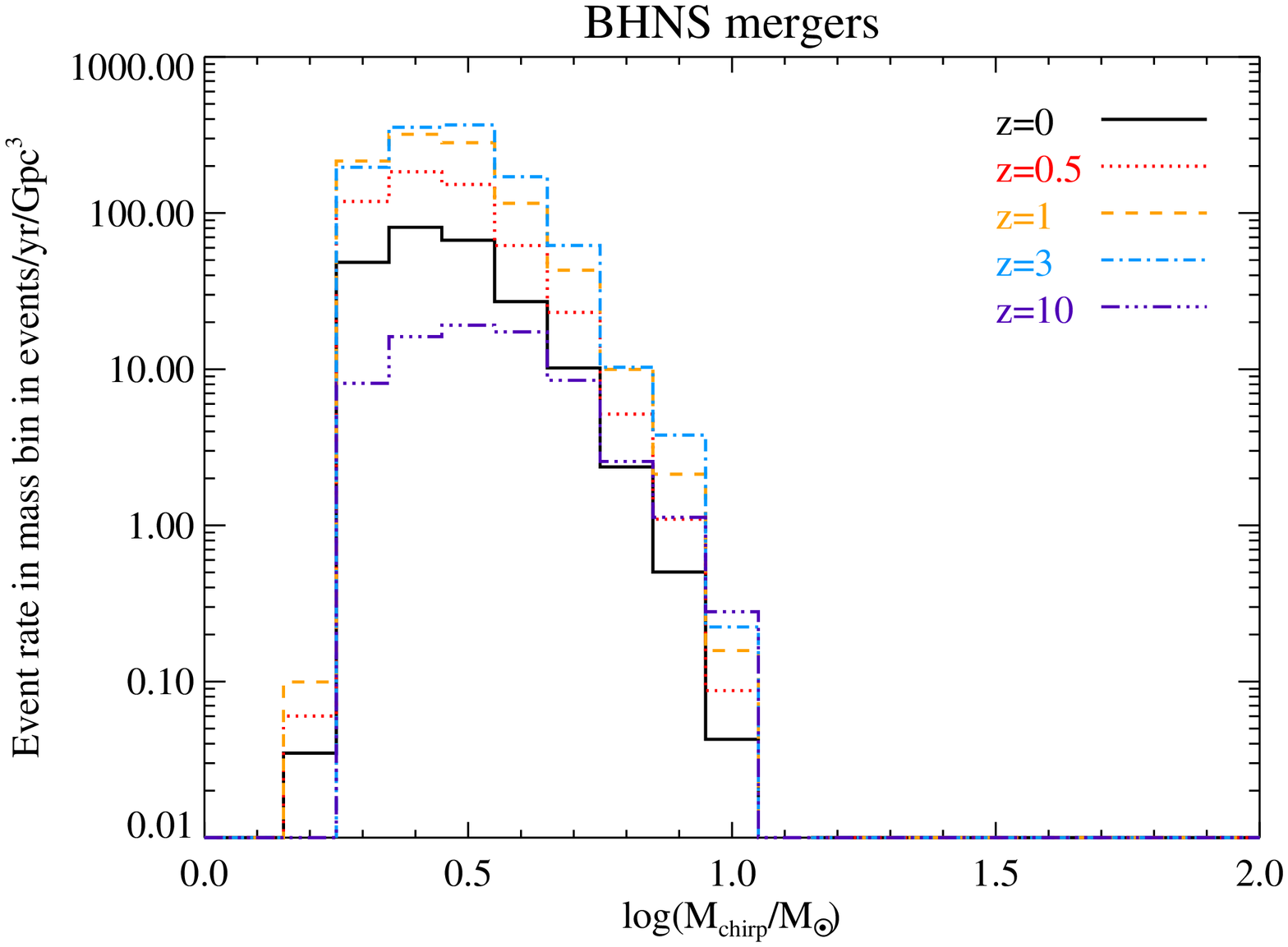}
\includegraphics[width=0.5\textwidth]{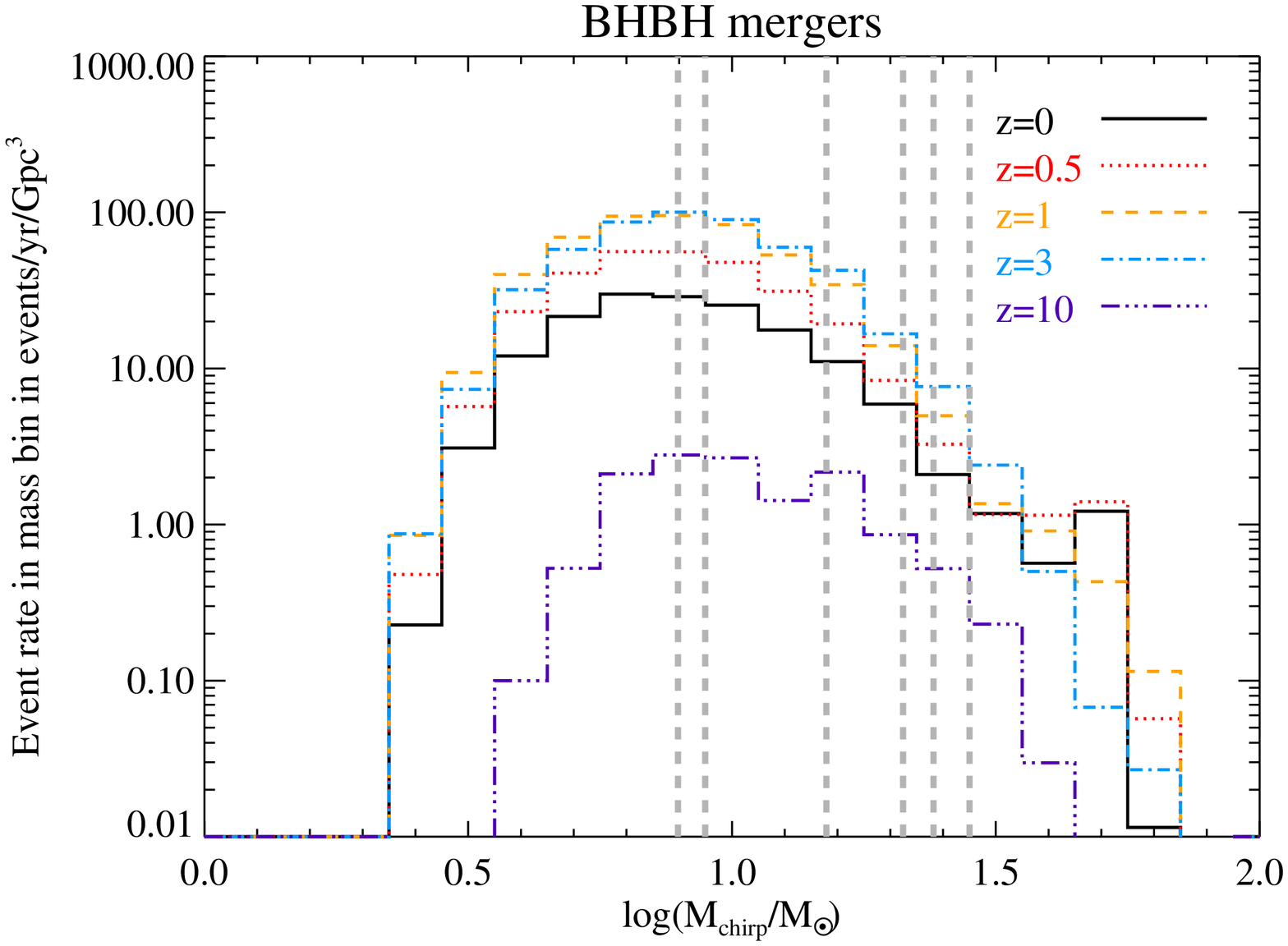}
\caption{The prediction of Chirp masses for the different merger types at various redshifts. \label{fig:rates3}. The vertical gray dash lines represents the currently detected GW events known.}
\end{figure}

A final prediction from our models involves the distribution of masses of the merging compact remnants that give rise to the observed GW events. We show these distributions in Figure \ref{fig:rates2} for NSNS, BHNS and BHBH mergers. The range of allowed Chirp masses (where $M_{\rm chirp}=(M_1 M_2)^{3/5}/(M_1+M_2)^{1/5}$) varies for each pair of merging remnants, due to the narrow range of possible NS masses (which we take to be between 1.4 to 3\,M$_{\odot}$) while black holes have an almost unrestricted mass range. 

We see that, surprisingly, we do not predict a significant change in the mass ranges until the highest redshifts, early in the age of the Universe when $z=10$. At lower redshifts the shape of the distribution varies little and only the total rate varies significantly.

When comparing our synthetic distributions to the GW events observed to date it must be remembered that here we are not taking account of the fact that higher Chirp mass events produce a stronger signal and are thus detectable to a greater distance. The distance to which events can be detected is $\propto M_{\rm chirp}^{5/6}$ and so the volume and total rate is $\propto M_{\rm chirp}^{5/2}$. This skews the distribution of the observed masses to higher chirp masses more in agreement with the observed event distribution.

\section{Discussion}\label{sec:disc}

\subsection{Caveats and uncertainties}

Our predictions habe uncertainties and caveats that must be appreciated and understood when evaluating their accuracy. These are the uncertain initial binary population parameters, the physics of our common envelope evolution prescription and the strength of the supernova kicks given to the compact remnants. We discuss how these parameters effect the our predicted rates for different events individually.

The initial parameters that describe the stellar populations we model, specifically the initial mass function and the primary-mass dependent binary fraction, initial period distribution and initial mass ratio distribution. While we have used a representative initial mass function the binary parameters are taken from  \citet{2017ApJS..230...15M} and are the current best available. Others have investigated how varying these numbers can affect the predictions of GW events \citep{2015ApJ...814...58D,2018MNRAS.474.2937C} and have found that at most the effect is a factor of 2 to 3 difference in predicted rates. We find similar changes  comparing the GW event rates from our v2.2 population in Table \ref{tab:rates} to those derived from our older BPASS v2.1 models in Table \ref{tab:rates2}, which used a different initial parameter distribution. By including more close binaries we have increased the BHBH merger rate by a factor of 2 while the NSNS and BHNS rates are nearly unchanged.

We note that we also now predict type Ia SN rates that are in line with observations, compared to our v2.1 results that predicted too many \citep{2017PASA...34...58E}. This change is due to our now including more single stars at lower masses where previously in v2.1 we had assumed all stars were in binaries. Changing the binary parameters however will not change the CCSN rate. To first order binary interactions do not have a significant impact on the overall supernova rate. This reveals the power of considering several transients at the same time in that their dependence on uncertainties is different.

The remaining uncertainties in our predictions are produced by the physics within our stellar models. These can be significant and order of magnitude in size as found by \citet{2015ApJ...814...58D}. The first aspect is the implementation of common envelope evolution within our stellar models. Our method is based on the conservation of energy between the binding energy of the envelope that is removed from the system and the orbital energy of the binary system \citep[see][for details]{2008MNRAS.384.1109E,2017PASA...34...58E}. We do not instantaneously remove the envelope, but do so at the fastest possible numerically stable rate. As a result we only have to assume a single parameter on the efficiency of the conversion of binding energy to orbital energy as in a detailed stellar evolution model we know the structure of the star and thus its binding energy accurately. 

In \citet{2017PASA...34...58E} we calculated the effective value of the $\alpha_{\rm CE}\lambda$, a constant representing the energy transfer efficiency, that our common envelope prescription provides.  We found values ranging from 100 for the smallest mass ratio systems up to of the order of 2 when the mass ratio is unity before CEE. Most CEE events have values in the range 2 to 30. These values are consistent with those others assume. We could vary the strength of the common envelope prescription in our models by introducing a multiplier to make the energy transfer more or less efficient. This would either increase or decrease the post-common envelope periods within our models. This would primarily effect the GW transient events rates and the type Ia double-degenerate channel rates but again leave the CCSN rate unchanged. 

We also note that the uncertainty in our common envelope prescription is degenerate to some degree with the uncertainty in the initial period distribution. For example, one way to get more close binaries to get more GW events is either to have more binaries with short initial periods or to have a less efficient energy transfer from binding energy to orbital energy. We suggest that given our results do agree generally with the different observed transient rates our common envelope prescription must give the correct answer to within an order of magnitude. Although further tuning of the model is likely to be required, this is numerically difficult as if we change the strength of our common envelope prescription we would need to recalculate our entire grid of stellar models which is beyond the scope of this study.

The third uncertainty to consider here is the strength of the kick the forming neutron stars and black holes are given in the supernova that forms them. This has been considered recently by \citep{2018MNRAS.474.2937C} and it shows that it is likely to be the most significant factor in determining the rate of NSNS GW events. This has also be confirmed by \citet{2018arXiv180404414B} who showed that by using a kick model related to the progenitor mass they were able to boost the NSNS merger rate by an order of magnitude.

To show how important the kick is we can compare the BPASS v2.2 results from Table \ref{tab:rates} to those from v2.2 involving the \citet{2018arXiv180404414B} in Table \ref{tab:rates2}. Comparing these numbers we see that while the NSBH and BHBH merger rates are only slightly affected the NSNS merger rate increases significantly. The near order of magnitude increase is significant and indicates that the supernova kicks are the primary source of uncertainty in predicted rates. As the known transient rates for compact objects become more accurately determined from future observing runs, the supernova kicks and other uncertain aspects will become more tightly constrained.

 \begin{table}
 \caption[]{Local Universe transient event rates, given the volume-averaged history of cosmic star formation and chemical enrichment as in Table \ref{tab:rates}. Here we list the GW event rates with a different initial parameter distribution (BPASS v2.1) and v2.2 models with the \citet{2018arXiv180404414B} kick model.}
    \label{tab:rates2}
    \begin{tabular}{lcc}
     \hline
      \noalign{\smallskip}
      Transient Type    &  log$_{10}$\,R($z=0$)$^a$  & log$_{10}$\,R($z=0.5$)$^a$  \\
      \noalign{\smallskip}
      \hline
      \noalign{\smallskip}
      v2.1 \\
      NSNS    &  2.61  $\pm$  0.02   & 2.89 $\pm$  0.02\\
      NSBH    &  2.35  $\pm$  0.03   & 2.63 $\pm$  0.01\\
      BHBH    &  1.80  $\pm$  0.05   & 1.96 $\pm$  0.04\\
\hline
\multicolumn{3}{l}{Bray \& Eldridge kick (v2.2)}\\
      NSNS    &  3.329  $\pm$  0.009   & 3.525 $\pm$  0.007\\
      NSBH    &  2.40  $\pm$  0.03   & 2.69 $\pm$  0.02\\
      BHBH    &  1.82  $\pm$  0.05   & 2.06 $\pm$  0.04\\
            \hline
          \end{tabular}\\

 { $^a$ Rates are given per type in units of events yr$^{-1}$\,$h_{0.7}^{3}$\,Gpc$^{-3}$}
  \end{table}

\subsection{Implications for Gravitational Wave Observatories}
 
For NS-NS merger transients, the mass range is relatively narrow. Nonetheless, the rate will be sensitive to the assumed metallicity history in a similar way; if more compact objects retain sufficient mass to become black holes, there may be a reduction in the relative rate of NS-NS mergers. Future observations of increased samples of short GRBs may constrain this through the occurrence rates of electromagnetic transients. However further complication in predicting gravitational wave transient rates is offered by the relatively small horizon of LIGO to these sources, which may be insufficient to average over the cosmic variance in the environs of the Local Group and nearby clusters. A more detailed analysis of large scale structure effects and the most probable host galaxies for such events will require a more detailed simulation informed by the patchy, merger-driven star formation and metal enrichment histories and spreads on galactic scales. Combining BPASS event rates with an N-body based, semi-analytic galaxy evolution model is likely the most promising route to such simulations.
 
The precise rates of observed GW events, and the mass distribution of their progenitors, will remain subject to small number uncertainties for some time due to the low rate of detectable events. The current upgrades to both LIGO and VIRGO, and potentially the addition of KAGRA and LIGO-India, will allow fainter triggers to be confidently identified, and so there will likely be improvement in the number statistics and in the sensitivity horizon (and thus volume probed). The rates of GW events are expected to vary by less than a factor of two between $z=0$ and $z=1$, and so the detailed redshift evolution (and hence metallicity dependence) will remain uncertain for some time to come, unless events can be localised to galaxies with well-constrained metallicity.

\subsection{Future constraints}

There are three key inputs to a physical analysis of transient rates: the intertwined cosmic star formation and metallicity histories, the stellar evolution models, and the observational selection effects.

The cosmic history of star formation and metal enrichment is being constantly refined, and independently determined using different samples and indicators. The rise of large area infrared surveys, including those obtained using the Herschel telescope, has allowed the hitherto hidden dust-obscured components of star formation to be added to those observed in visible and ultraviolet light. The history of galaxy formation is also being modelled through ever more precise N-body and hydrodynamical simulations. Uncertainties do remain. There is likely to be considerable variation on galactic scales, and in cluster environments. Nonetheless, the volume-averaged results are relatively robust and will be further refined by future observations with the James Webb Space Telescope (JWST) and other large scale facilities.

The corrections applied to determine volumetric rate estimates of transients from observed number counts are also moderately well understood. Current uncertainties on observational event rates are dominated by corrections for the electromagnetic transient luminosity function and, in the case of beamed sources, the jet opening angle. Survey sensitivity and completeness corrections can introduce systematic offsets, which suggests that a uniform survey is desirable, but these are now being modelled using sophisticated simulation and the scatter on rate estimates from different surveys has decreased over time.

The Large Synoptic Survey Telescope (LSST) is expected to begin observations in 2022. The 8\,m class telescope is expected to identify and provide redshifts for of order one hundred thousand of type Ia supernova to $z=1.2$ through their photometric colours and light curves, with the primary goal of constraining dark energy. The number of CCSNe detected will likely be several times higher \citep{2009arXiv0912.0201L,2017arXiv170804058L,2018RPPh...81f6901Z}. These will simultaneously provide a uniform and precise measurement of the volumetric density of these events which will allow the redshift evolution of the supernova population to be measured with a precision hitherto impossible. Between $z=1$ and $z=0$, we expect the core collapse supernova rate to evolve by a factor of six, while the type Ia rate declines by a much smaller factor of two. Thus reducing their relative volumetric rate uncertainties to $<10\%$ (requiring samples of $\sim$1000 events in each class per redshift bin) would allow testing of this theoretical scenario and any alternatives as necessary.

In samples this size, statistical corrections for observational selection will dominate rate uncertainties and here a single, uniform survey has obvious benefits. However the precision of LSST-derived rates will suffer from the limited plans for rapid spectroscopic follow up of these relatively faint transients. If a significant fraction of events is followed up spectroscopically, and classified, it will also be possible to contrast the volumetric rates in Type II (II-L, II-P, II-n), Type Ib and Type Ic supernovae - a sensitive probe of the mass loss and stripping of massive stars in binary systems. LSST's deep, high-cadence supernova survey may also facilitate identification of rare transient sources such as PISNe, and may provide an alternate route to identify some compact mergers through kilonova emission, although this is strongly dependent on the final survey cadence selected.

Rate estimates for GRBs (both long and short) will likely remain uncertain for some time. Most of the current constraints arise from the Neil Gehrels Swift Observatory, due to its sensitivity and ability to localise a large fraction gamma-ray transients. The Fermi Gamma-Ray Space Telescope and INTErnational Gamma-Ray Astrophysics Laboratory (INTEGRAL) also continue to provide burst alerts, with different energy sensitivity ranges. While this makes them potentially more sensitive to short (hard) GRBs than Swift, both suffer from poor burst localisation and limited follow-up, such that very few bursts have redshifts.  It is possible that the new generation of ground-based, wide field images designed for GW transient follow-up \citep[such as the Gravitational Optical Transient Observatory, GOTO,][]{2017NatAs...1..741S} will be able to localise a larger fraction of future events, but it is not clear whether these will be priorised.

The Space-based multi-band astronomical Variable Objects Monitor mission \citep[SVOM, due to launch in 2021,][]{2016arXiv161006892W} will produce a GRB sample smaller than that accumulated by current Swift observations, but incorporates systematic rapid follow up, and aims to substantially improve number counts for distant ($z>5$) bursts, which are strongly sensitive to the metallicity enrichment history within the first Gyr of cosmic history. In this sense, it will be complementary to improved measurements of star formation rate variation and metal enrichment at these epochs with the JWST telescope. SVOM also aims to have increased sensitivity to the hard spectra of short bursts, their thermalised prompt emission and their time evolution, when compared to Swift. Hence, it is expected to provide constraints on NS-NS and potentially NS-BH mergers.

The more ambitious, and expensive, Theseus mission \citep[currently being considered for ESA's M5 mission to launch in 2032,][]{2018AdSpR..62..191A} will address the question in more detail. Theseus, while a longer term prospect, expects to detect GRBs at an event rate an order of magnitude higher than the existing Swift telescope, meaning that it will equal the precision of existing GRB rate estimates within one year, and far exceed it over the mission lifetime. It will also greatly improve constraints on the GRB luminosity function. while existing instruments such as Fermi and Integral are already probing events with energies outside of the peak sensitivity range of Swift.

So, given these sources of information, is there potential for transient rates to constrain the third input: the physics of stellar evolution models, particularly at the low metallicities and in the binary environments required to create the most massive transients?

\citet{2018MNRAS.477.4685B} suggested that future GW event detections will be able to highly constrain the input physics of binary population synthesis codes. However they ignored the fact that we also must understand the metallicity and star-formation evolution of the Universe. Here by using the same model codes to predict the rate of SNe we have a second complimentary dataset that is mostly sensitive to the star-formation history of the Universe. The CCSN rate effectively directly measures the star-formation rate while the type Ia SN provides a measurement of older star-formation history if we can reproduce the delay-time distribution of these events. The fact here we can reproduce the observed rates and delay-time distributions for both SNe types means that if our predicted GW event rates do not match constraints from future observations then it must be our stellar models that are incorrect.

The relevant uncertain aspects of the stellar evolution models are the mass-loss rates, common envelope evolution and the neutron-star/black-hole kicks. These all have a minor effect on the overall supernova rate and primarily will change the SNe to be of different types observationally. However, as discussed above, they all have an effect on the rate of GW events as they determine how close two compact remnants are when they explode, and thus how long they will take to merge via gravitational radiation. Only by such a multi-event type approach as we use here is it possible to start to test aspects of the stellar evolution physics.

BPASS is a detailed stellar evolution code in which each individual stellar model takes several minutes to run and $\sim250,000$ models feed into the population synthesis analysis required to produce figure \ref{fig:rates}. The computing resources required to examine a grid of such models with varying stellar astrophysics is only today becoming available. However simulations based on the rapid population synthesis codes, e.g. BSE and COMPAS (which uses semi-analytic formalisms to estimate the evolution of stars with physical properties between fixed grid points), have already suggest that reasonable changes to the input stellar physics can cause up to an order of magnitude variation in compact binary merger rates and that stellar physics constraints are obtainable \citep[e.g.][]{2018arXiv180608365T,2018MNRAS.474.2959G}. Given the ability of binary stellar population synthesis codes, such as BPASS, to model multiple flavours of transients, the increased precision in rate density measurements which should be obtained from both LSST and LIGO may be sufficient to start to constrain different physical assumptions. This will be explored in future work using BPASS models.





\section{Conclusions}\label{sec:conc}

Our main conclusions can be summarised as follows:
   \begin{enumerate}
      \item We combine event rates predicted by BPASS population synthesis models as a function of age and metallicity with analytic prescriptions for the cosmic volume-averaged star formation rate and chemical evolution history. 
      \item We determine the redshift dependence of transient event rates for both electromagnetic and gravitational wave transients from the same stellar population synthesis model.
      \item These are in excellent agreement with current observational constraints, and with the rate estimates determined from gravitational wave sources, albeit with uncertainties on the observed source luminosity functions and the opening angle of relativistically-beamed emission.
   \end{enumerate}
   
We aim to continue using BPASS to constrain the joint evolution of different astrophysical transients as data statistics improve. We also plan to refine our models. In particular, we aim to analyse the effect of  variations in the analytic star formation and metallicity histories in this study, and also to replace volume averaged rates with metallicity spread and star formation environment as determined from cosmological models. Ultimately, we will use this to look at transient rates in cluster versus field environments and as a function of galaxy/cluster mass, as well as exploring the effects of stellar physics and multiplicity statistics.

\section*{acknowledgements}
   JJE and ERS acknowledge travel funding and support from the University of Auckland. ERS also acknowledges support from the Australian Astronomical Observatory through the Shaw Distinguished Visitor scheme. We thank Joe Lyman for useful conversations.
   BPASS is enabled by the resources of the NeSI Pan Cluster. New Zealand's national facilities are provided by the NZ eScience Infrastructure and funded jointly by NeSI's collaborator institutions and through the Ministry of Business, Innovation \& Employment's Research Infrastructure programme. URL: https://www.nesi.org.nz.

 \bibliographystyle{mnras} 
 \bibliography{events} 

%
%

\bsp	
\label{lastpage}
\end{document}